

\magnification = 1200
\hsize = 15 truecm
\vsize = 23 truecm
\baselineskip 20 truept
\voffset = -0.5 truecm
\parindent = 1 cm
\overfullrule = 0pt
\count 0=0
\footline={\hfil}
\def\u{1\kern -0.7em \hbox {1}}
\def\p{\partial}

\settabs 3 \columns
\+&&  KUL-TF-94/7 \cr
\+&&  hep-th/9404011 \cr
\+&&  March 1994 \cr

\null

\centerline
{\bf Topological conformal field theory  with a rational $W$ potential}
\centerline
{\bf  and  the dispersionless KP hierarchy}

\vskip 1 truecm

\centerline{\ S. Aoyama$^1$}

\smallskip
\centerline{\ Instituut voor Theoretische Fysica}
\smallskip
\centerline{\ Katholieke Universiteit Leuven}
\smallskip
\centerline{\ Celestijnenlaan 200D}
\smallskip
\centerline{\ B-3001 Leuven, Belgium}

\vskip 0.5cm

\centerline{\ Y. Kodama$^2$}
\smallskip
\centerline{\ Department of Mathematics}
\smallskip
\centerline{\ Ohio State University}
\smallskip
\centerline{\ Columbus, Ohio  43210,  USA}

\vskip  0.5cm

\noindent
{\bf Abstract}

We present a new class of topological conformal field theories (TCFT)
characterized by a rational $W$ potential, which includes the minimal models
of A and D types as its subclasses.
An explicit form of the $W$ potential is  found by solving the underlying
dispersionless KP hierarchy in a particular small phase space.
 We discuss also the dispersionless KP hierarchy in large phase spaces by
 reformulating the hierarchy, and show that the $W$ potential takes a universal
form, which does not depend on a specific form of the solution in a large
space.

\vskip 1 cm
\noindent $^1$ e-mail : shogo\%tf\%fys@cc3.kuleuven.ac.be

\noindent $^2$ e-mail : kodama@math.ohio-state.edu

\vfill\eject

\footline={\hss\tenrm\folio\hss}

\noindent
1.~~~~  Since Witten's foundation topological field theories have undergone
 intensive investigations. Among them TCFT's $^{[1]}$ are most characteristic.
Namely it was shown in ref. 2 that
genus-zero 3-point functions $c_{ijk} = <\phi_i\phi_j\phi_k>$ of BRST
invariant primaries $\phi_i$ are such that
$$
\eqalignno{
c_{0ij} & = {\rm const.}, \quad ({\rm flatness\ of\ metric}),  & (1a) \cr
\partial_m c_{ijk} & = \partial_i c_{mjk}, \quad
({\rm integrability}),  & (1b)  \cr
c_{ij}^{\ \ m} c_{mkl} & = c_{jk}^{\ \ m}c_{mli}, \quad
({\rm associativity}).    & (1c)   \cr}
$$
Through the Landau-Ginzburg formulation they solved these equations by
the A-D-E- minimal models and elucidated the intriguing connections of TCFT's
 with matrix models, dispersionless KP hierarchy, singularity theory,
 etc.. This pioneering work stimulated a
number of people to study eqs. (1a)$\sim$(1c) and to find other
types of TCFT$^{[3,4]}$. As had already been shown in ref 5, TCFT's thus
 obtained
may be coupled with $2-d$ gravity.  The studies in refs. 6, 7 and 8
showed that the Landau-Ginzburg formulation is also a suitable framework
 for this generalization.

In the Landau-Ginzburg picture TCFT's are completely determined by the
$W$ potential. All the solutions to eqs. (1a)$\sim$(1c) so far found in the
literature$^{[3]}$
give the $W$ potential in a polynomial form of several variables.
However the potential of the D-model is exceptional, as expressed in
a non-polynomial form by eliminating one of the variables $^{[2]}$.
 There barely appears a non-polynomial piece as deforming  the $W$ potential
 of the A model$^{[6,7,8]}$.
A  rational $W$ potential of a more general form has been proposed in the
recent work 9 for a characterization of a certain class of multi-matrix models.
However, they did not discuss a connection to  TCFT.
A non-polynomial $W$ potential was also studied for
a Landau-Ginzburg description of the $c=1$ string in ref. 10. But it is not
rational either and does not describe a TCFT.

 In this note we show that the $W$ potential in ref. 9  gives new types of TCFT
 satisfying eqs. (1a)$\sim$(1c). The novelty of these TCFT's is the presence of
a finite number of positive and negative primaries.  They contain the A-D-
models as subclasses of this model.
An explicit form of the $W$ potential is found by
properly solving the underlying dispersionless KP equation in a particular
phase space.
We here give a practical formula to evaluate genus-zero 3-point functions
$c_{ijk}$ in terms of the $W$ potential, and show that the model satisfies eqs.
(1a)$\sim$(1c).
Both positive and negative flows also appear in the dispersionless Toda
hierarchy $^{[11]}$, and
 Takasaki's extension$^{[12]}$ of the dispersionless BKP hierarchy, where they
discussed
the general structure of these  hierarchies.
Our model is  however considered as a reduction of the dispersionless
KP hierarchy extended by adding negative flows, rather than as the
dispersionless BKP hierarchy $^{[12]}$. A further extension of the
dispersionless KP hierarchy including several types of flows has
been discussed in ref. 4. A tau-function (free energy) is found, and is used to
construct
 correlation functions. But it remained to be  rather a formal expression.

To discuss a topological Landau-Ginzburg theory coupled with $2-d$
gravity, it is indispensible to extend the analysis in the small phase
space to  that in larger ones. So far little has been  understood in
this case even for the A
model. Following ref $^{[13]}$ we reformulate the
dispersionless KP theory in the framework of quasi-linear system of partial
differential equations. The reformulation turns out to be  optimal
to discuss  on the issue. We are able to show a universal form of
the $W$ potential in the entire phase space.

\vskip 1cm

\noindent
2.~~~~  We consider the dispersionless KP hierarchy
$$
{\partial W \over \p t_i} = \{Q_i, W \}
                       = {\p Q_i \over \p p}{\p W \over \p t_0}
                        - {\p Q_i \over \p t_0}{\p W \over \p p}, \quad  {\rm
for} -\infty < i < \infty , \eqno (2)
$$
with the $W$ potential in a rational form${[9]}$
$$
\eqalign{
W = {1 \over n+1}p^{n+1} & + v_{n-1} p^{n-1} +  \cdots\cdots + v_0   \cr
  &  + {w_1 \over p-s} + \cdots\cdots + {w_{m-1} \over (m-1)(p-s)^{m-1}} +
   {w_{m} \over m(p-s)^m} .  \cr}  \eqno (3)
$$
where $v_i$ and $w_i$ are the functions of $t_j,\  -\infty < j  < \infty$.  The
Hamiltonian functions $Q_i$ for $t_i$ flows are defined by
$$
\eqalign{
Q_i & = [{1 \over i+1}\lambda^{i+1}]_+,
                \quad \quad \quad {\rm for} \quad  \quad i \ge 0,     \cr
Q_{-1} & =  \log (p-s),       \cr
Q_{-i} & =    [-{1 \over i-1}\mu^{i-1}]_- ,
                 \quad \quad \quad {\rm for} \quad  \quad i \ge 2 ,    \cr }
                 \eqno (4)
$$
where $\lambda$ and $\mu$ are the semi-classical limits of the Lax operators
given by
$$
\eqalign{
\lambda & = [(n+1)W]^{{1 \over n+1}} = p + O({1 \over p}),
\quad {\rm for \ \ large}\quad p ,  \cr
\mu     & = [mW]^{{1 \over m}} = {\root m \of {w_m} \over p-s} + O(1).
\quad {\rm for \ \ small }\quad  p-s.  \cr} \eqno (5)
$$
Here $[..]_{+}$ and $[..]_{-}$ indicate the parts of non-negative and negative
powers in $p$, respectively.
Eqs. (5) imply $W ={{ \lambda ^{m+1}} \over m+1} ={{ \mu ^m} \over m}$
 which may be considered as a Riemann-Hilbert problem
 on a genus zero surface (sphere) defined by the potential (3)
 $^{[12]}$.
The compatibility among the flows in (2), i.e. ${\p \over \p t_i}{\p \over
\p t_j} W = {\p \over \p t_j}{\p \over \p t_i} W$, can be shown by
the zero curvature conditions,
$$
{\p Q_i \over \p t_j} - {\p Q_j \over \p t_i} + \{Q_i, Q_j\} = 0, \quad
\quad {\rm for} -\infty < i, j < \infty,
$$
which can be directly derived from (4).
In ref. 9 the dispersionless KP hierarchy of this kind was discussed as
characterizing a certain class of matrix models.
 But their hierarchy did not contain the negative $t_{-i}$ flows
 with $i \ge 2$. Eq. (2) can not then give a $W$ potential of TCFT
 except for the case $m=1$, as it will be clear in this note.

We now define the fields $\phi_i$ as  Laurent polynomials of $p$,
$$
\phi_i = {d Q_i \over dp}, \quad \quad {\rm for} -\infty < i <
\infty.   \eqno (6)
$$
For an arbitrary integer $a$, a set of $(n+m+1)$ fields
$$
\phi_a,\ \  \phi_{a+1},\ \cdots\cdots\cdots, \ \ \phi_{a+n+m-1},\ \
\phi_{a+n+m},
$$
form a basis of a Laurent polynomial ring with the ideal ${d W\over d p} (=W')
= 0$.
When $a=-(m+1)$ they are primaries, satisfying the flatness condition as shown
below.
The fusion algebra is found by calculating as
$$
\phi_i \phi_j = c_{ij}^{\ \ l}\phi_l + W'Q_{ij},\quad -(m+1) \le i,j \le n-1.
\eqno (7)
$$
We propose to write 3-point functions  as
$$
<\phi_i\phi_j\phi_k > = -\oint_C [{\phi_i\phi_j\phi_k \over W'}],
\quad -\infty < i,j,k < \infty, \eqno (8)
$$
in which $C$ is a contour surrounding roots of $W'= 0$. Note here that if $W$
is a polynomial in $p$, i.e. the $A$ model, the contour integral can be
evaluated as the residue at $p=\infty$.  Note also that $W'$ is nilpotent in
the numerator of the integrand, and the 3-point functions are faithful to the
fusion algebra. Consequently they satisfy the associativity given by eq. (1c).
Let $\eta_{ij}$ be the metric defined as
$$
<\phi_i\phi_j\phi_0> = \eta_{ij},  \quad \quad -(m+1) \le i,j \le n-1.
\eqno (9)
$$
Then from eqs. (7) and (8) it follows that
$$
<\phi_i\phi_j\phi_k > = c_{ij}^{\ \ l}\eta_{lk},
\quad \quad -(m+1) \le i,j,k \le n-1.   \eqno (10)
$$
We denote all other $\phi_i$'s with $i \le -(m+2)$ or $ n \le i$ by
$$
\eqalign{
\sigma_N (\phi_i) & = c_{N,i} \phi_{N(n+1)+i},
\quad {\rm for}\quad 0 \le i \le n-1,  \cr
\sigma_N (\phi_{-i}) & = d_{N,i} \phi_{-(Nm+i)},
\quad {\rm for} \quad 2 \le i \le m+1,
\cr}   \eqno (11)
$$
where
$$
\eqalign{
c_{N,i} & = [(i+1)(i+1+n+1)\cdots\cdots(i+1+(N-1)(n+1))]^{-1},  \cr
d_{N,i} & = [(i+1)(i+1+m)\cdots\cdots(i+1+(N-1)m)]^{-1},    \cr}
$$
and $N \ge 1$.
They are identified  as the descendants for the primaries$^{[6,7]}$. But
note that the primary $\phi_{-1} (= \sigma_0 (\phi_{-1})) $ has no descendant.
This is typical in the Toda hierarchy.
Note also that $\sigma_N (\phi_n) = \sigma_N (\phi_{-(m+1)})$ for all $N$'s by
$W'=0$.

\vskip 1 cm

\noindent
3.~~~~  We now study the disperionless KP hierarchy (2) in a small phase space
by restricting the flow parameters $t_i$ only to those with $ -(m+1) \le i \le
n-1$.
Let us assume that the potential $W$ satisfies
$$
{\p W \over \p t_0} = 1.      \eqno (12)
$$
which guarantees the flatness of the metric as shown below.
Then eq. (2) becomes
$$
{\p W \over \p t_i} = \phi_i,      \eqno (13)
$$
for $-(m+1) \le i \le n-1$, from which one can easily see  separation of the
variables,
$$
\eqalignno{
s & = t_{-(m+1)} ,      & (14a)  \cr
{\p \over \p t_{-i}} v_j & = 0,
\quad \quad {\rm for}\quad 1 \le i \le m+1, \quad 0 \le j \le n-1,  & (14b) \cr
{\p \over \p t_i} w_j  & = 0,
\quad \quad {\rm for}\quad 0 \le i \le n-1, \quad 1 \le j \le m,  & (14c) \cr}
$$
Note that eq. (14a) is the unique solution for $s$.
Furthermore from eq. (13) together with (6) and (7) it follows that
$$
{\p \phi_j \over \p t_i} = Q_{ij}', \quad \quad -(m+1) \le i,j \le n-1 .
\eqno (15)
$$
For $ij \le 0$ , this equation is evident since both sides vanish due to eqs.
(14a)$\sim$ (14c).
If $i \ge 0$ and $j \ge 0$
it can be shown by noting that
$$
Q_{ij} = [{\phi_i \phi_j \over W'}]_+
= [\phi_i \lambda^{j-n}]_+.
$$
If $i<0$ and $j<0$ the proof goes in the same way.

In refs. 8 and 14 eq. (15) is refered as the flatness condition of the metric,
and leads to the flatness of the metric defined by the period integral. We
shall show that it is also a sufficient condition for the flatness of the
metric defined by
(9) with the residue formula (8).
There is a gap between these two metrics. The flatness of our model may be
shown by studying the relation between the residue formula and the period
integral. Following ref. 8 we assume
  1-point functions of the fields $\phi_i$ by the  period integral: for the
primaries
$$
\eqalign{
<\phi_i> & = {1 \over (i+n+2)(i+1)} \oint_{p=\infty} dp\ \lambda^{i+n+2},\cr
& \quad \quad \quad \quad \quad \quad \quad \quad \quad
\quad \quad \quad \quad \quad \quad \quad
 \quad 0 \le i \le n-1,  \cr
<\phi_{-i}> & = -{1 \over (i+m-1)(i-1)} \oint_{p=s} dp\ \mu^{i+m-1}, \cr
& \quad \quad \quad \quad \quad \quad \quad \quad \quad
\quad \quad \quad \quad \quad \quad \quad
 \quad 2 \le i \le m+1,  \cr} \eqno (16)
$$
and for all other fields
$$
\eqalign{
\sigma_N (\phi_i) & = c_{N+2,i} \oint_{p=\infty} dp \ W^{N+1+ {i+1 \over n+1}},
 \quad \quad  0 \le i \le n-1,  \cr
\sigma_N (\phi_{-i}) &  =  d_{N+2,i} \oint_{p=s} dp \
W^{N+1+ {i-1 \over m}}, \quad \quad  2 \le i \le m+1, \cr}     \eqno (17)
$$
with  $c_{N+1,i}$ and
$d_{N+1,i}$ given in eq. (11).
The period integral fails to define a 1-point function of the primary
$\phi_{-1}$, since it diverges by a naive extension of the formula (16) to this
case.  An explicit form of $<\phi_{-1}>$ will be later given.

Let $<\psi>$ be one of these 1-point functions. Then we can show
the following relation with the 3-point function defined by eq. (8):
$$
<\phi_i \phi_j \psi > =
{\p \over \p t_i}{\p \over \p t_j} <\psi>.  \eqno (18)
$$
A similar formula appeared for the $A$ model. But having the different contours
on both  sides eq. (18) is not obvious for the generalized model. The easiest
way$^{[8,14]}$ to show this may be to differentiate $\psi$ twice  by the flow
parameters and find  the Gauss-Manin system
$$
{\p \over \p t_i}{\p \over \p t_j}<\psi> =
c_{ij}^{\ \ l }{\p \over \p t_l } {\p \over \p t_0 } <\psi>,
\quad {\rm for} \quad  -(m+1) \le i,j, \le n-1,  \eqno (19)
$$
by eq. (15).
The same equation can be derived by using eq. (7) in the l.h.s. of eq. (18).
This proves  eq. (18) indirectly. We here give a direct proof.
Consider the case when $\psi = \phi_k, \ 0 \le k < \infty$. By eqs. (13) and
(16) the r.h.s. of eq. (18) may be calculated as
$$
\eqalign{
{\p \over \p t_i}{\p \over \p t_j}<\phi_k> &  =
\oint_{p=\infty} dp\ { \phi_i\phi_j\phi_k \over W'} \cr
& + \ \oint_{p=\infty} dp\ {\phi_i \phi_j {d \over dp }[{1 \over
k+1}\lambda^{k+1}]_- \over W'} \ + \ \oint_{p=\infty} dp\ [{1 \over
k+1}\lambda^{k+1}]Q'_{ij}.  \cr}  \eqno (20)
$$
If $-(m+1) \le i,j \le -1$, the second piece in the r.h.s is vanishing, while
the remaining pieces add up to give
$$
{\p \over \p t_i}{\p \over \p t_j}<\phi_k> = c_{ij}^{\ \ l}
\oint_{p=\infty} dp\ { \phi_l\phi_k \over W'},
$$
by eq. (7).
Since this integral has no residue at $p=s$, we can analytically deform the
contour around $p=\infty$ to the one surrounding the roots of $W'=0$. This
leads to (18),
$$
{\p \over \p t_i}{\p \over \p t_j}<\phi_k> = -c_{ij}^{\ \ l}
\oint_C dp\ { \phi_l\phi_k \over W'}
= -\oint_C dp\ { \phi_i\phi_j\phi_k \over W'} .
$$
If $i$ or $j$ takes other values, the second and third integrals in eq. (20)
are either cancelled with each other by eq. (7) or trivially vanishing. The
first integral has no residue at $p=s$, so that it becomes
$<\phi_i\phi_j\phi_k>$ by the analitical deformation of the contour. For the
case when $\psi = \phi_{-k}, \quad 0 < k < \infty$,
the same proof goes through.

The meaning of eq. (18) is twofold. On one hand it provides us with a practical
 way to evaluate the residue of eq. (8). Without this, eq. (8) would be a
formal definition.
On the other hand eq. (15) guarantees the integrability
$$
<\phi_i \phi_j > = {\p \over \p t_i}<\phi_j > = {\p \over \p t_j}<\phi_i >,
  \eqno (21)
$$
for $-(m+1) \le i,j \le -2, \quad 0 \le i, j \le n-1$.
It is not evident a priopri , for instance , in the case when $ij < 0$ .
Thus we obtain the unique higher point functions by differentiating the
the lowest ones given by eqs. (16) and (17).
We were not able to define a 1-point function of $\phi_{-1}$ by the period
integral. But it can be found by  solving eq. (18) with $\psi = \phi_{-1}$, or
equivalently eq. (21) with $i$ or $j = -1$:
$$
<\phi_{-1}> = \int^s_0 dp \ [W]_+ \ - \ {1 \over m}\oint_{p=s} dp \ W
\log[(p-s)^m W]   + f(t_{-1}).   \eqno (22)
$$
Here $(p-s)^m$ works as a regulator without which the second integral diverges.
The function $f(t_{-1})$ is to be fixed by
$$
<\phi_{-1}\phi_{-1}\phi_{-1}> = \partial_{-1}^2 <\phi_{-1}>.
$$

Equipped with eq. (22) the formula (18) is now valid for $\psi = \phi_{-1}$ as
well. As the result we find the property of  3-point functions $c_{ijk}$ given
by eq. (1b). The flatness of the metric (9) can be also shown
by means of this formula.  For $0 \le j \le n-1$ we calculate as
$$
\eqalign{
{\p \eta_{ij} \over \p t_k}  & =
{\p \over \p t_k}{\p \over \p t_i}{\p \over \p t_0}<\phi_j>  \cr
  & = (j-n)c_{ki}^{\ \ l} \oint_{p=\infty} dp \
\lambda^{j-2n-1}\phi_l = 0.   \cr}
$$
For $-(m+1) \le j \le 0$ the same thing can be shown.

\vskip 1 cm

By eqs. (8), (17) and (18) we can show the recursion relation for the
descendants $\sigma_N (\phi_i)$ with $i\not= -1$
$$
<\sigma_N (\phi_i)\phi_j\phi_k> = <\sigma_{N-1}
(\phi_i)\phi_l><\phi^l\phi_j\phi_k>,
$$
which is characteristic for TCFT's coupled with $2-d$ gravity$^{[5,7]}$.
 For a derivation it suffices to note the formulae
$$
\eqalign{
<\sigma_N (\phi_i ) \phi_0 > & =  <\sigma_{N-1} (\phi_i)>, \cr
\sigma_N (\phi_i) & = <\sigma_N (\phi_i) \phi_0 \phi^l>\phi_l\ + \ W'Q,   \cr}
$$
with a suitable Laurent polynomial $Q$.

\vskip 1 cm

\noindent
4.~~~~  Let us now solve eq. (13) to find an explicit form of the $W$
potential. By eqs. (4) and  (6) it turns into the equations for $w_i$ and $v_i$
$$
\eqalign{
{\p \over \p t_{-k}}{\p \over \p t_{-l}} w_i & =
(i-1){\p \over \p t_{-(k+l-m-1)}}w_{i-1}, \quad \quad 1 \le i \le m, \cr
{\p \over \p t_k}{\p \over \p t_l} v_i & = (i+1){\p \over \p t_{k+l-n}}v_{i+1},
\quad \quad  0 \le i \le n-1, \cr}
$$
respectively.
By solving them recursively we obtain
$$
\eqalign{
w_i & = \sum_{l_1 + \cdots + l_i = (i-1)m + i}
t_{-l_1}t_{-l_2}\cdots\cdots t_{-l_i},  \cr
 \ \ \cr
v_i & = t_i + \sum_{j=2}^{n-i-1}
{(i+j-1)! \over {j (i!)}}
  \sum_{ l_1+\cdots +l_j  = \atop (j-1)n+i+j-1}
 t_{l_1}\cdots\cdots t_{l_j}, \cr
s & =  t_{-(m+1)}. \cr
 \ \ \cr}  \eqno (23)
$$
The $W$ potential with these solutions gives a TCFT, for which  3-point
functions  are calculated by the residue formula (8) or  eq. (18).
As has been remarked, the primary $\phi_{-1}$ has no desendant in this TCFT.
 Its 1-point function was not given by the period integral, but eq. (22)
which contains logarithmic pieces $\propto \log t_{-m}$ . These odd phenomena
regarding $\phi_{-1}$ are the consequences of the presence of the $t_{-1}$ flow
which is characteristic  for the Toda hierarchy and the requirement of the
fusion algebra (10), or equivalently the associativity  (1c).
If we look at  subrings, the $W$ potential with eqs. (23) gives a subclass of
TCFT's, which does not have $\phi_{-1}$.
The simplest one is the A model which has only positive primaries $\phi_i, \  0
\le i \le  n-1$ $^{[2]}$. It is given by the $W$ potential (3) in which all the
negative flow parameters are switched off, $w_i=0$ for all $i$. Or we may
consider the case where $n$ and $m$ are multiples of an integer $M (\ge 2)$,
i.e., $n = Ma$ and $m = Mb$.
The primaries
$$\phi_{-Mb},\ \phi_{-M(b-1)},\ \cdots\cdots,\ \phi_{M(a-2)},\ \phi_{M(a-1)},
 \eqno (24)
$$
form a subring. The $W$ potential of the TCFT with these primaries is given by
eq. (23), where $t_i = 0$ with $i$ non-multiple of $M$  (e.g. $s = 0$).
The primary $\phi_{-1}$ is moded out, and hence the odd phenomena due to
$\phi_{-1}$ disappear. As a result, all the primaries (24) have the descendants
defined according to eq. (11). For both fields the 1-point functions are given
by the period integral of the $W$ potential. They  are evaluted to be
polynomials of the flow parameters $t_i$ corresponding to the primaries (24).
The D model is the special case  of this subclass of TCFT with $M=2$ and $b=1$
$^{[2]}$.
Thus the TCFT given by the $W$ potential with eq. (23) is a natural
generalization of the A-D- models.

\vskip 1cm

\noindent
5.~~~~  So far eq. (2) was discussed in the small phase space. We may be
naturally interested in solutions in larger spaces. The previous arguments can
be generalized
in a universal way.
To do this we reformulate eq. (2) following the works in ref. 13.
First of all we invert  eqs. (5) in terms of the local coordinates $\lambda$
and $\mu$:
$$
\eqalign{
p & = \lambda - {u_{n-1} \over \lambda} - {u_{n-2} \over \lambda^2} -
 \cdots\cdots\cdots -  {u_0 \over \lambda^n} - O({1 \over \lambda^{n+1}}), \cr
p & = s + {u_{-m} \over \mu} + {u_{-(m-1)} \over \mu^2} +
 \cdots\cdots\cdots +  {u_0 \over \mu^{m+1}} + O({1 \over \mu^{m+2}}). \cr}
$$
Note here that the coefficients in the higher orders are expressed in terms of
$u_i$'s.
For convenience, we write $u_i$ and $\phi_i,\ -(m+1) \le i \le n-1$, as column
vectors
$$
\eqalign{
U & = [u_{n-1},u_{n-2}, \cdots\cdots \cdots,u_{-m},u_{-(m+1)}]^T, \cr
\Phi & = [\phi_{n-1},\phi_{n-2}, \cdots\cdots
\cdots,\phi_{-m},\phi_{-(m+1)}]^T, \cr}
$$
with $s = u_{-(m+1)}$.
There are recursion relations between these quantities which can be expressed
as
$$
\Phi^T (A - p\ \u) = [1,0,\cdots\cdots \cdots,0,0](\phi_{-(m+1)} - \phi_n),
$$
where $A$ is an $(n+m+1)\times (n+m+1)$ matrix given by

$$
\left(\, \matrix{
0        &  1\hfill   &\ldots\ldots   & 0  & 0  & 0  & 0  &\ldots\ldots   & 0 &
0     \cr
-u_{n-1} & 0 \hfill  &   &   & 0  & 0  &   &   &  & 0    \cr
\vdots   &  \vdots\hfill &   &  &\vdots   &\vdots   &   &   &  & \vdots     \cr
\vdots   &  \vdots\hfill &   &   &\vdots   &\vdots   &   &   &   &\vdots
\cr
-u_2     &    &\ldots\ldots  & 0  & 1  &
&   &   &  &        \cr
-u_1     &   -u_2\ \ \hfill  &\ldots\ldots   & -u_{n-1}  &  0 &
1   & \hfill 0\hfill  &\dots\ldots   & 0 &  0        \cr
0 & 0 \hfill &\ldots\ldots  & 0 & 0  &
\ \ s\ \ & \ \ u_{-m} & \ldots\ldots &\ \  u_{-2} & u_{-1}  \cr
0  &  &  &  & 0  &
0 & s & \ldots \dots &  & u_{-2} \cr
\vdots  &  &  &  & \vdots &
\vdots &  &  & \vdots & \vdots  \cr
\vdots  &  &  &  & \vdots &
\vdots &  &  & \vdots & \vdots  \cr
0 &  &  &   & 0 &
0 &  &  & s & u_{-m} \cr
1 &  0\hfill  &\ldots\ldots  & 0 & 0 &
0 &  0  &\ldots\ldots  &  0 & s   \cr} \right).
$$

\vskip 0.5cm

\noindent
These recursion relations lead to a closed form for $\phi_i,\ -(m+1) \le i \le
n-1$:

$$
\phi_{-i} = {1 \over (p-s)^i} \det\left|\,\matrix{
u_{-m}&u_{-(m-1)}\hfill&\ldots\ldots&u_{-(m-i+3)}\hfill&u_{-(m-i+2)}\hfill \cr
& \ & \ & \ & \ & \cr
-(p-s)&\ \ u_{-m}\hfill&\ldots\ldots&u_{-(m-i+4)}\hfill&u_{-(m-i+3)}\hfill\cr
& \ & \ & \ & \ & \cr
\hfill\, \vdots\hfill&\,
\ \ \vdots\hfill&&\,  \ \ \ \vdots\hfill&\, \ \ \vdots\hfill \cr
\hfill\, \vdots\hfill&\,
\ \ \vdots\hfill&&\,  \ \ \ \vdots\hfill&\, \ \ \vdots\hfill \cr
& \ & \ & \ & \ & \cr
0&\ \ 0\hfill&\ldots\ldots&\ \ \ u_{-m}\hfill & u_{-(m-1)}\hfill \cr
0&\ \ 0\hfill&\ldots\ldots&-(p-s)\hfill&\ \ u_{-m}\hfill \cr}\right|,
$$
for  $2 \le i \le (m+1)$,
$$
\phi_i  = \det\left|\,\matrix{
p&-1\hfill&\ldots\ldots&\hfill 0\hfill&\hfill\ \ 0\hfill \cr
& \ & \ & \ & \ & \cr
u_{n-1}&\ \ p\hfill&\ldots\ldots&\hfill 0\hfill&\hfill\ \  0\hfill\cr
& \ & \ & \ & \ & \cr
\hfill\,   \vdots\hfill&\,
\ \ \vdots\hfill&&\,  \ \ \vdots\hfill&\, \ \ \vdots\hfill \cr
\hfill\,  \vdots\hfill&\,
\ \ \vdots\hfill&&\,  \ \ \vdots\hfill&\, \ \ \vdots\hfill \cr
& \ & \ & \ & \ & \cr
u_{n-i+2}\hfill&u_{n-i+3}&\ldots\ldots& p\ \ & -1\hfill \cr
& \ & \ & \ & \ & \cr
u_{n-i+1}&u_{n-i+2} &\ldots\ldots&u_{n-1}\hfill&\ \ p\hfill \cr}\right|,
$$
for $ 2 \le i \le n$, and
$$
\phi_{-1} = {1 \over p-s},\quad \phi_0 = 1, \quad \phi_1 = p.
$$
By noting
$$
W' = \phi_n -  \phi_{-(m+1)},
$$
it can be shown that eq. (2) is equivalent to a set of the following equations
$$
\eqalignno{
{\p U \over \p t_i} & = \phi_i (A) {\p U \over \p t_0}, \quad \quad
-\infty < i < \infty,  & (25)\cr
{\p W \over \p u_i} & = \phi_i, \quad \quad -(m+1) \le i \le n-1. & (26) \cr}
$$
Here the quantities $\phi_i (A)$ are $(n+m+1) \times (n+m+1)$ matrices given by
(6) with $p$ substituted by $A$.
The solution of (26) has been already found, i.e.,  eq. (23) in which $t_i$ are
replaced by $u_i$. With this  we
obtain a universal form of the $W$ potential in terms of $u_i$.
Thus the dispersionless KP hierarchy (2) is reduced to the quasi-linear system
(25).
Note that the variables $u_i$'s are the conserved densities for the system
(25), and they can be expressed by the period integrals as well as the free
energy
 as a consequence of the zero curvature conditions of $Q_i$'s, or
equivalently the compatibility conditions of the flows in (2)$^{[12,13]}$.

In order to construct some of the exact solutions of (25), we first note the
following: Due to the Cayley-Hamilton theorem, any field
 $\phi_M (A),\  M \le -(m+2),\  n+1 \le M$, can be decomposed into the
primaries,
$$
\phi_M (A)  = \sum_{-(m+1) \le i \le n-1} \Delta_i (U)\phi_i (A),  \eqno (27)
$$
with appropriate coefficients $\Delta_i (U)$. This is equivalent to
saying that $\phi_i$'s give a basis of the finite ring of Laurent
polynomials by $W'=0$.  For instance, we have for $M=n+1$
$$
\phi_{n+1}(A) = \sum_{-(m+1) \le i \le n-1} u_i\phi_i (A). \eqno (28)
$$
Putting eqs. (25) and (27) together gives
$$
{\p U \over \p t_M} = \sum_{-(m+1) \le i \le n-1} \Delta_i (U) {\p U \over \p
t_i}.  \eqno (29)
$$
We have infinitely many equations of this sort.
They  constrain solutions of the dispersionless KP hierarchy (2), and might be
related to the Virasoro constraints. In ref. 13 it was shown that from eq. (29)
 a solution of eq (25) in the small phase space with $t_M = 1$ can be
constructed as a hodograph form,
$$
\Delta_i (U)  = t_i,  \eqno (30)
$$
The explicit forms of $u_i$'s are then
obtained by inverting the algebraic equations (30).
 In particular, for $M=n+1$ we find the flat solution, i.e. $u_i=t_i$ with
$t_{n+1}=1$.
Of course  we may be interested in more general solutions of eq. (25) with (29)
in large spaces. An important point of the reformulation (25) and (26) is that
one can bring any solution $W$ of eq. (2) in  the universal form by finding an
appropriate solution
 $u_i,\ -(m+1) \le i \le n-1$ of eq. (25).
Dependence on the flow parameters $t_i,\ -\infty < i < \infty$, appears  only
implicitly through the solution $u_i$.

\vskip 1cm

\noindent
6.~~~~  In this note we have studied the dispersionless KP hierarchy (2) with
the rational  $W$ potential (3). In the small phase space it was solved by eqs.
(23). We have shown that this solution gives a TCFT, for which  the 3-point
function $c_{ijk}$  was given by the residue formula (8). The novelty of this
TCFT is the presence of positive and negative primaries. We have given the
proofs of the flatness of the metric $c_{0ij}$ and the integrability
$\partial_m c_{ijk}  = \partial_i c_{mjk}$ in some details, since they were not
evident by simply generalizing the arguments for the A model. The key step in
the proofs was to write the residue formula (8) in terms of the $W$ potential,
i.e., eq. (18). The TCFT thus obtained contains the A-D- models as subclasses
of TCFT.

We have also discussed the dispersionless KP hierarchy (2) in the entire phase
space. Through the reformulation
the arguments in the small phase spaces were universally extended to larger
ones. We have shown that any solution $W$ of the dispersionless KP hierarchy
(2)
in the entire phase space can be brought into  the universal form (23) with the
flow parameters $t_i$ replaced by an appropriate solution $u_i,\ -(m+1) \le i
\le n-1$, of eq. (25).

\vfill\eject

\noindent
{\bf Acknowledgments}

S. A. would like to thank the Research Council of K.U. Leuven for the financial
support.
The work of Y. K. was partially supported by NSF Grant DMS-9109041.

\vskip 2cm

\noindent
{\bf References}

\item{1.} T. Eguchi and S.-K. Yang, Mod. Phy. Lett. A5(1990)1693.
\item{2.} R. Dijkgraaf, H. Verlinde and E. Verlinde, Nucl. Phys. B352(1991)59.
\item{3.} B. Blok and A. Varchenko, Intern. J. Mod. Phys. A7(1992)1467;
\item{} E. Verlinde and N.P. Warner, Phys. Lett. B269(1991)96;
\item{} I. M. Krichever, Comm. Math. Phys. 143(1992) 415;
\item{} A.C. Cadavid and S. Ferrara, Phys. Lett. B267(1991)193;
\item{} S. Cecotti and C. Vafa, Nucl. Phys. B367(1991)359;
\item{} W. Lerche, D.-J. Smit and N.P. Warner, Nucl. Phys. B372(1992)87;
\item{} Z. Maassarani, Phys. Lett. B273(1991)457;
\item{} A. Klemm, M.G. Schmidt and S. Theisen, Int. J. Mod. Phys.
A7\-(1992)\-6215;
\item{} B. Dubrovin, Comm. Math. Phys. 145(1992)195; Nucl. Phys.
B379\-(1992)627;
``Differential geometry of the space of orbits of a Coxeter group"
 hep-th/9303152;
\item{} W. Lerche, ``Generalized Drinfeld-Sokolov hierarchies, quantum rings
and $W$ gravity", CERN-TH.6988/93, hep-th/9312188;
\item{} B. Zuber, ``On Dubrovin topological field theories", SPht 93/147,
hep-th/9312209.
\item{4.} I.M. Krichever, ``The $\tau$-function of the universal Whitham
hierarchy, matrix models and topological field theories", LPTENS-92/18,
hep-th/9205110.
\item{5.} E. Witten, Nucl. Phys. B340(1990)281;
\item{} R. Dijkgraaf and E. Witten, Nucl. Phys. B342(1990)486.
\item{6.} A. Losev, ``Decendants constructed from matter field in topological
Landau-Ginzburg theories coupled to topological gravity", hep-th/9211089;
\item{} A. Losev and L. Polyubin, ``On connection between topological
Landau-Ginzburg gravity and integrable systems", hep-th/9305079.
\item{7.} T. Eguchi, H. Kanno, Y. Yamada and S.-K. Yang, Phys. Lett.
B305\-(1993)\-235.
\item{8.} T. Eguchi, Y. Yamada and S.-K. Yang, Mod. Phys. Lett. A8(1993)1627.
\item{9.} L.Bonora and C.S. Xiong, ``Correlation functions of two-matrix
models", SISSA-ISAS 172/93/EP, BONN-HE-45/93, hep-th/9311089.
\item{10.} D. Ghoshal and S. Mukhi, ``Topological Landau-Ginzburg model of
two-dimen\-sional string the\-ory", MRI-\-PHY\-/13\-/93, TIFR\-/TH/\-93-62, \
hep\--th\-\-/9312189;
\item{} A. Hanany, Y. Oz, M. Plesser, ``Topological Landau-Ginzburg formulation
and integrable structure of 2d string theory", IASSNS-HEP-94/1, TAUP-2130-93,
WIS-93/123/Dec-PH, hep-th/9401030.
\item{11.} K. Taksaki and T. Takebe, Lett. Math. Phys. 23 (1991) 205.
\item{12.} K. Takasaki, Lett. Math. Phys. 29(1993)111.
\item{13.} Y. Kodama and J. Gibbons, Phys. Lett. A135(1989)167;
``Integrability of the dispersionless KP hierarchy", in: Proc. Workshop
``Non-linear Processes in Physics" (World Scientific, 1990) p. 166;
\item{} Y. Kodama, Phys. Lett. A147(1990)477.
\item{14.} N.P. Warner, ``$N=2$ supersymmetric integrable models and
topological field theories", USC-93/001, hep-th/9301088.

\bye